\documentclass[a4paper]{article}

\usepackage{graphicx}
\usepackage[utf8]{inputenc}
\usepackage[T1]{fontenc}
\usepackage[english]{babel}
\usepackage{amsmath}
\usepackage{amsfonts}
\usepackage{multirow}

\def\beq{\begin{equation}}
\def\eeq{\end{equation}}
\def\bsp#1\esp{\begin{split}#1\end{split}}

\newcommand\epem {e^+e^-}
\newcommand\kt {k_{\bot}}
\newcommand\yc {y_{\mathrm{cut}}}
\newcommand\Ec {E_{\mathrm{cut}}}
\newcommand\dd {\mathrm{d}}
\newcommand\ordo[1] {\mathcal{O}(#1)}

\begin{document}


\thispagestyle{empty}

\begin{flushright}
MITP/19-072
\end{flushright}

\vspace{1.5cm}

\begin{center}
{\LARGE \bf Jet transition values for the anti-$\kt$ algorithm}\\
\vspace{1cm}
{\large Zolt\'an Sz\H{o}r}\\
\vspace{2mm}
      {\small \em PRISMA Cluster of Excellence, Institut f{\"u}r Physik, }\\
      {\small \em Johannes Gutenberg-Universit{\"a}t Mainz,}\\
      {\small \em D - 55099 Mainz, Germany}\\
\end{center}

\vspace{1.5cm}

\begin{abstract}\noindent 
We define jet transition values for the anti-$\kt$ algorithm for both hadron and $\epem$ colliders.
We show how these transition values can be computed and how they can be used to improve 
the performance of clusterization when jet resolution parameters are varied over a larger set of values.
Finally we present a simple performance test to illustrate the behavior of the new 
method compared to the original one.
\end{abstract}

\vspace*{\fill}

\newpage


\section{Introduction}
\label{sec:introduction}
The production of hadronic jets is a common feature of particle collisions.
Jets are widely studied, as they can be used to test the standard model
and measure its parameters, they can signal new physics, and provide important background 
for new physics searches as well.

Jets are defined through jet clustering algorithms:
they take final state particles as an input and combine them according to their prescription 
into larger objects, what we then call jets.
The algorithms have a set of resolution parameters, which defines the jet structure:  
fixing the values of jet algorithm parameters determines what happens in each step of the 
clusterization, what particles get combined into jets eventually.
Although jet algorithms are required in all kind of jet analysis, one particularly 
important observable is the so-called jet rate.  
The jet rate, as a function of its parameters, directly connects to the 
clustering algorithm, as it provides useful information about how the number of jets depends 
on the choice of parameters.

Jet rate measures the relative production rate of n-jets compared to all hadronic events.
It is given by the ratio of the n-jet cross section $\sigma_{\mathrm{n-jet}}$ 
and the total hadronic cross section $\sigma_{\mathrm{tot}}$ at center-of-mass energy $Q^2$:
\begin{equation}
\label{eq:Rn}
R_n(\vec{a}) = \frac{\sigma_{\mathrm{n-jet}}(\vec{a})}{\sigma_{\mathrm{tot}}}\,,
\end{equation}
where $\vec{a}$ denotes the set of jet resolution parameters characteristic to a given jet algorithm.
Jet rates are mostly studied as a function of one or more of their resolution parameters.
This means clustering the same set of momenta with a wide range of chosen values of 
the jet resolution parameters. 
This set of momenta might represent a point in the phase space of final state particles 
or a physical event, but the actual representation is not important in the scope of the paper. 
Thereby we use the umbrella term 'partonic event'.

Since repeated clusterization is usually computationally inefficient, in practice 
one tries to exploit the properties of the algorithm to enhance performance in computations.
This is also important on the theory side, since making higher order predictions in perturbation 
theory typically requires the generation of millions of phase space points, which all need to be clustered 
individually.
Although the bulk of computational cost is coming from the calculation of amplitudes and subtraction terms, 
slow clusterization might add a non-negligible time contribution as well.

In the case of $\epem$ colliders the most common jet clustering algorithm is the $\kt$ (or Durham) 
algorithm \cite{Catani:1991hj}, which has auspicious properties to do computations efficiently, 
illustrated in the next section.
Today, in the LHC era, the commonly used jet algorithm is the anti-$\kt$ algorithm \cite{Cacciari:2008gp}.
Though jet rate studies similar to $\kt$ ones are not prevalent, they are also hampered by the lack 
of properties that the $\kt$ has.
Hence computations can be slowed down significantly due to clusterization only.

In this paper we present a reformulation of the anti-$\kt$ algorithm equivalent to the original, 
which makes it possible to define transition values in a similar fashion to the $\kt$ algorithm.
Furthermore we show how these transition values can be computed and used to speed up calculations.
Our method can be used for both hadron and $\epem$ colliders.


\section{The $\kt$ algorithm}
\label{sec:kt}
We start with a short review of the $\kt$ algorithm, and discuss how it is used in calculations 
in practice.
The algorithm depends on a single jet resolution parameter $\yc$ and the distance measure is defined as
\begin{equation}
y_{ij} = \frac{2 \mathrm{min}(E_i^2,E_j^2) (1- \cos \theta_{ij})}{Q^2} \,.
\end{equation}
$E_i$ and $E_j$ denote the energy of particle $i$ and $j$ respectively, while $\theta_{ij}$ labels 
the angle between the three-momenta $\vec{p}_i$ and $\vec{p}_j$.
During clusterization we compute $y_{ij}$ for each pair of particles and find the smallest one 
$y_{kl}=\mathrm{min} \, y_{ij}$.
If $y_{kl} < \yc$ holds we combine particles $k$ and $l$, then start the procedure again with the new 
list of objects.
Otherwise we stop the clusterization and the resulting objects are considered jets.

In the case of the $\kt$ algorithm one can uniquely define transition values.
Transition values $y_{i-1 \leftarrow i}$ are certain values of $\yc$, where the number of jets changes 
from $i$ into $i-1$ for a given final state configuration.
The distribution of the transition value behaves as an event shape observable.
Using the $\kt$ algorithm every transition value $y_{i-1 \leftarrow i}$ can be computed performing
the clusterization only once independently of $\yc$, 
such that in every clusterization step the smallest $y_{kl}$ value provides the corresponding $y_{i-1 \leftarrow i}$ 
transition value. 
We repeat the steps until all particles are clustered into two jets.
When jets are defined through the $\kt$ algorithm the number of jets is a monotonically decreasing 
function of $\yc$ for every possible partonic event.

These two properties of the algorithm described previously make possible to connect the 
$\mathrm{d} \sigma/\mathrm{d} y_{i-1 \leftarrow i}$ differential distributions 
and the $\sigma_{\mathrm{n-jet}}(\yc)$ cross section.
For example the three-jet cross section can be computed as
\begin{equation}
\label{eq:3jetdist}
\sigma_{3-jet}(\yc) = \int_{\yc}^1 \dd y_{2 \leftarrow 3} \frac{\dd \sigma}{\dd y_{2 \leftarrow 3}}
- \int_{\yc}^1 \dd y_{3 \leftarrow 4} \frac{\dd \sigma}{\dd y_{3 \leftarrow 4}} \,.
\end{equation}
The meaning of the two terms are the following:
the three-jet cross section for a chosen $\yc$ gets contributions from the $\dd \sigma /\dd y_{2 \leftarrow 3}$
differential cross section for every $y_{2 \leftarrow 3}$ value which is greater than $\yc$.
This gives the first term in Eq. (\ref{eq:3jetdist}).
However the resulting quantity in itself would include all events with $y_{3 \leftarrow 4} \in [0,1]$.
Events with $y_{3 \leftarrow 4} \in [0,\yc]$ are indeed events which cluster into 
three-jets, however for events with $y_{3 \leftarrow 4} \in [\yc,1]$ clustering  
stops at four-jets. 
Thus we need to subtract the integrated $\dd \sigma/\dd y_{3 \leftarrow 4}$ 
distribution to get the correct three-jet cross section, which gives the second term.
Eq. (\ref{eq:3jetdist}) provides a very useful relation to speed up numerical calculations. 
One has to perform the clusterization only once per partonic event, calculate the differential cross sections, 
then do a simple integration with the desired $\yc$ according to the formula to obtain the n-jet cross section.


\section{The anti-$\kt$ algorithm}
\label{sec:antikt}
Now we turn our interest towards the anti-$\kt$ algorithm and discuss its 
shortcomings in computational time compared to the $\kt$ algorithm.
The anti-$\kt$ algorithm uses two different measures: a two-particle measure $d_{ij}$ and 
a beam jet measure $d_{iB}$.
They are defined as
\begin{equation}
\begin{split}
d_{ij} & = \mathrm{min}(k_{\bot, i}^{2p},k_{\bot, j}^{2p}) 
\frac{(y_i - y_j)^2 + (\phi_i - \phi_j)^2}{R^2} \,, \\
d_{iB} & = k_{\bot, i}^{2p} \,,
\end{split}
\end{equation}
for hadron colliders, where $k_{\bot, i}$, $y_i$ and $\phi_i$ denote 
the transverse momentum, rapidity and azimuth of particle $i$ respectively.
In the case of $\epem$ colliders we have
\begin{equation}
\begin{split}
d_{ij} & = \mathrm{min}(E_i^{2p},E_j^{2p}) \frac{(1-\cos \theta_{ij})}{1 - \cos R} \,, \\
d_{iB} & = E_i^{2p} \,,
\end{split}
\end{equation}
with the notation being identical to the one introduced in the previous section. 
Choosing $p=-1,0,1$ we obtain the anti-$\kt$ \cite{Cacciari:2008gp}, the Cambridge/Aachen \cite{Dokshitzer:1997in} 
and the inclusive $\kt$ \cite{Catani:1991hj} algorithms respectively.
Collectively they are named as the general inclusive $\kt$ algorithm.

The anti-$\kt$ algorithm has two jet resolution parameters: $R$ and $\Ec$.
During clustering we calculate $d_{iB}$ for every particle $i$ and $d_{ij}$ for every particle pair $i,j$.
If $d_{kl}$ is the smallest measure, we combine particle $k,l$, 
but if $d_{kB}$ is the smallest one, particle $k$ is considered a jet candidate, and we remove it
from the list of objects.
We repeat these steps until every particle becomes part of a jet candidate.
Finally we apply energy cut(s), and every jet candidate with $E_i > \Ec$ is a resolved jet.

The anti-$\kt$ algorithm has characteristics and properties, which makes it 
preferable for experimental use \cite{Cacciari:2008gp}, for example cone-like jet shapes.
However the algorithm has certain other properties, which unfortunately make computational shortcuts 
like Eq. (\ref{eq:3jetdist}) absent, therefore making clusterization more expensive 
in the study of the jet rate observable. 
This is due to the fact that in general the number of jets is not a monotonic function of 
$R^2$ or $1-\cos R$ as it can be seen in Fig. \ref{fig:njets}.
The reason is partially the presence of the additional $\Ec$ parameter. 
Although we obtain more and more jet candidates when we increase the spatial resolution, 
many of them would not survive the last cut on the energy.
Furthermore the presence of the beam jet measure, $d_{iB}$ prevents the same definition of 
jet transition values as in the case of the $\kt$ algorithm.

This would leave us in an unfortunate situation where clustering would need to be done for 
each different choice of the jet resolution parameters, in particular when we vary $R$.
We note that this still can be an issue, even if one uses the the improved version of the 
anti-$\kt$ algorithm \cite{Cacciari:2005hq}, which scales $\ordo{N \log N}$ compared 
to the $\ordo{N^3}$ cost of the original formulation.
The choice of the efficient method depends on the number of histogram bins and the number of 
partons to be clustered.

Fortunately we can still define jet transition values, which can be used in calculations.


\section{Transition values}
\label{sec:transvalues}
We start with an equivalent reformulation of the anti-$\kt$ algorithm, which is more suitable 
to define and find transition values.
First we combine the two measures $d_{ij}$ and $d_{iB}$ the following way
\begin{equation}
\label{eq:defyijk}
y_{ijk} \equiv \yc \frac{\mathrm{min}_{i,j} \, d_{ij}}{\mathrm{min}_k \, d_{kB}} \,, 
\end{equation}
where we define $\yc \equiv R^2$ and $\yc \equiv 1 - \cos R$ for hadron and $\epem$ colliders respectively.
Note that $y_{ijk}$ is independent of $\yc$.

Now clusterization is done as it follows: first we calculate $y_{ijk}$. 
If $y_{ijk} < \yc$, we combine particle $i,j$; otherwise we consider particle $k$ to be a jet candidate 
and remove it from the list.
We repeat the procedure until the list is empty.
Finally we apply the energy cuts on our jet candidates.

The clustering procedure is now similar to the $\kt$ algorithm, hence we can define jet transition values 
in a similar fashion.
We call $y_{t} \equiv \yc$ a transition value when the clustered particle configuration changes.
It is important to notice that it does not necessarily imply a change in the number of jet candidates. 
Two different $\yc$ values can result in the same number of jet candidates, but these candidates 
may differ in their momenta configuration. 
As an illustration let us consider the following: we have 4 partons 
such that they can be separated into two hemispheres and we cluster them into 3 jets. 
The parton in the first hemisphere is the hardest and is widely separated in angle from the other three, 
we label it by $H$. 
In the second hemisphere one parton is soft and two are hard, labeled as $s$, $h$ and $h'$ respectively, with 
the following angle separation $1 - \cos \theta_{s h} \sim 1 - \cos \theta_{s h'} > 1 - \cos \theta_{h h'}$. 
If we choose $\yc$ such that $1 - \cos \theta_{s h} > \yc > 1 - \cos \theta_{h h'}$, then in the clustering process 
we first remove the soft parton $s$ from the list, then combine partons $h$ and $h'$ together and finally obtain 
$(s),\, (h h'),\, (H)$ as jet candidates. 
In an other case with $\yc > 1 - \cos \theta_{s h}$ we first combine partons $s$ and $h$, and obtain 
$(sh), \, (h'),\, (H)$ as jet candidates. 
Both configurations have the same number of jet candidates, but the way each parton is associated to a jet is different, 
hence they are in two different regions separated by a transition value. 
This behavior is due to the presence of the beam jet measure $d_{iB}$. 
Then the final number of resolved jets depends on the chosen value of $\Ec$ as well.

Using this definition is convenient in practice. 
The transition values must be calculated only once, 
then one can apply as many different energy cuts as wanted without repeating the clusterization again.
Nevertheless the calculation of $y_{t}$ values is not straightforward.
For the $\kt$ algorithm the sequence of clustering is independent of $\yc$ and relevant information can be fully 
retrieved for any $\yc$ value from one complete clusterization.
In contrast, the clusterization sequence of the anti-$\kt$ algorithm depends on the actual choice 
of $\yc$, due to the presence of the two different distance measures.

It was shown that in the Cambridge algorithm one faces a similar problem, but transition values can still be
found systematically \cite{Bentvelsen:1998ug}.
Here we can adopt the method of Ref. \cite{Bentvelsen:1998ug} as well to find transition values for the 
anti-$\kt$ algorithm in the following way:
\begin{enumerate}
\item First set an initial value for $y_{\mathrm{ini}}$ and set $\yc = y_{\mathrm{ini}}$.
\item If $\yc$ is less than some preset lower limit $y_{\mathrm{stop}}$, stop the algorithm.
\item Perform clusterization with the chosen $\yc$, and find the maximum value of $y_{ijk}$ during the process.
\item Store the transition value $y_{t} = y_{ijk}^{max}$ and apply energy cuts to obtain the corresponding 
number of jets.
\item Set $\yc = y_{ijk}^{max}$ and go to Step 2.
\end{enumerate}

Clusterization between two transition values is completely determined, choosing two different 
$\yc$ in this set will lead to the same jet configuration.
This leads to an improvement in speed in the calculation of jet rates.
We can fill histograms more easily between two transition values, we do not have to consider 
each bin separately and perform clusterization over and over again.

It is worth to mention that the method is independent of the definition of $d_{ij}$ and $d_{iB}$
and also independent of how $d_{ij}$ and $d_{iB}$ are calculated, given that the maximum value of 
$y_{ijk}$ can be obtained during clusterization. 
Therefore the transition method can be used both in the hadron and $\epem$ collider version of 
the anti-$\kt$ algorithm and in fact for any version of the general inclusive $\kt$ algorithm, that being 
the original $\ordo{N^3}$ or the improved $\ordo{N \log N}$ version. 
\begin{figure*}[ht]
\centering
\includegraphics[width=0.75\textwidth]{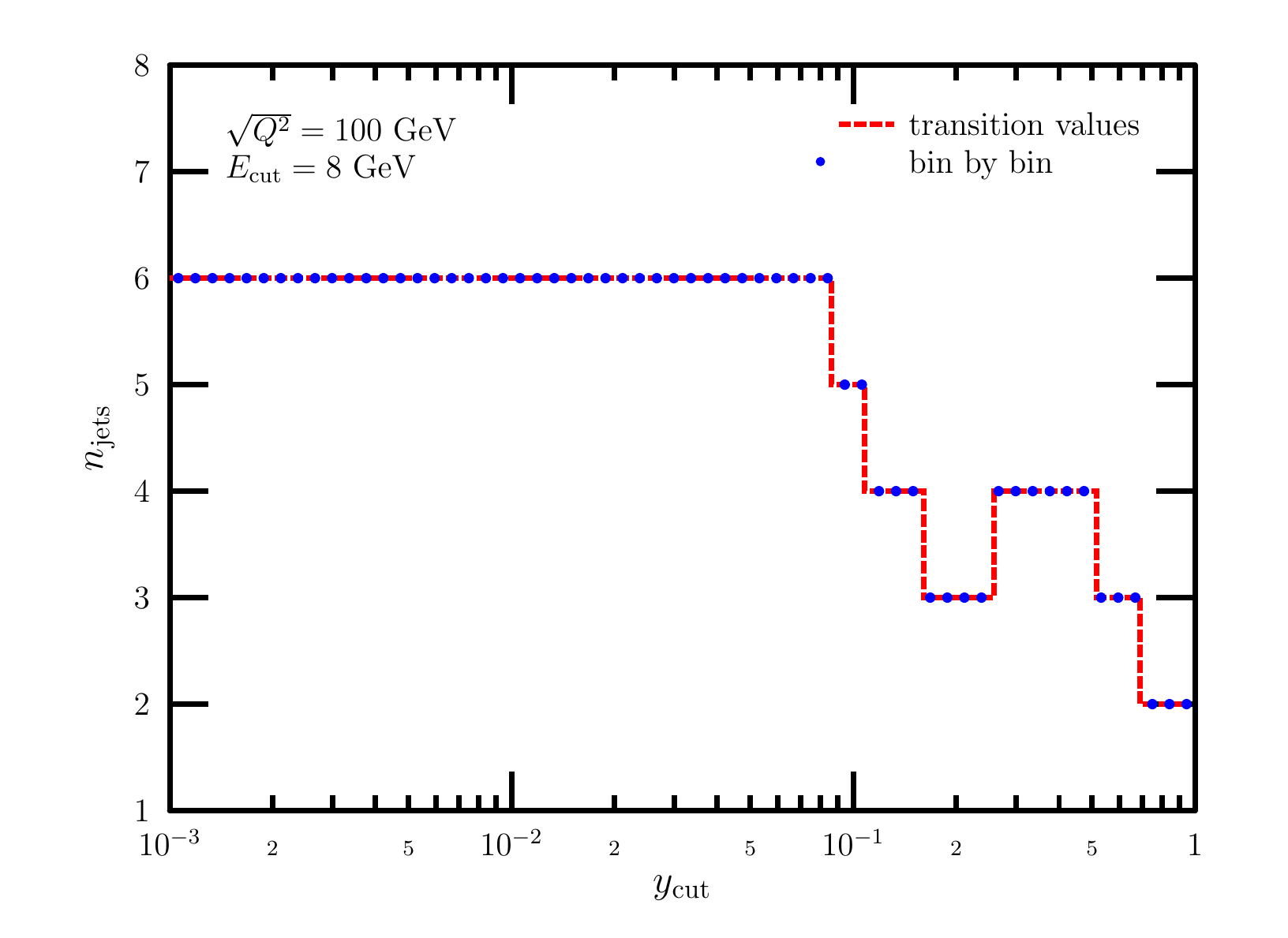}
\caption{The number of jets as function of $\yc$ obtained from clustering a randomly generated partonic event 
with 10 particles in two different ways. The two approaches provide identical results and the non-monotonic 
behavior of function is also visible.}
\label{fig:njets}
\end{figure*}

On Fig. \ref{fig:njets} we show the number of jets as a function of $\yc$. 
We used a randomly generated partonic event with 10 particles in the final state at 
$\sqrt{Q^2} = 100$ GeV center-of-mass energy.
$\Ec$ was chosen $8$ GeV.
The 10 particle configuration was clustered with the $\epem$ version of the anti-$\kt$ algorithm using 
both approaches: bin-by-bin with 30 $\yc$ values denoted by blue dots and via the transition values 
method denoted by the red line.
Both methods produce identical results, but while the bin-by-bin method required 30 repeated full clusterizations,
the red curve was reproduced from 14 transition values.
Fig. \ref{fig:njets} also illustrates the general non-monotonic behavior of the number of jets as a function 
of $\yc$.


\section{Performance}
\label{sec:performance}
Finally we explore the performance of the new method compared to the traditional
approach.
We employ three different methods: 
we name the method computing the number of jets over a wide range of $\yc$ through transition values 
as \textit{transition}, while the bin-by-bin version is dubbed as \textit{direct}.
Both the \textit{transition} and the \textit{direct} method are based on the original formulation of 
anti-$\kt$, which scales as $\ordo{N^3}$.
As third we include the \texttt{FJcore} version of the anti-$\kt$ algorithm from the \texttt{FastJet} 
package \cite{Cacciari:2011ma}, and perform bin-by-bin clustering with it.  
We call this method \textit{fjcore}.
We note that the \texttt{FJcore} package provides only a $\ordo{N^2}$ scaling in contrast to the full 
\texttt{FastJet} version scaling as $\ordo{N \log N}$. 
In exchange \texttt{FJcore} is easier to integrate and is more likely to be used in cases, where 
the whole apparatus of \texttt{FastJet} is not required. 
We implemented the \textit{transition} and the \textit{direct} methods in a \texttt{Fortran90} program 
and included the \texttt{FJcore} algorithm through the provided wrapper. 
For simplicity we chose the $\epem$ collider version of the anti-$\kt$ algorithm.
Using \texttt{RAMBO} \cite{Kleiss:1985gy} we generated 1000 partonic events with 5, 10, 15 and 20 particles 
in the final state, and clustered them with all three methods.
We checked that the three methods give the same results, as it is illustrated in Fig. \ref{fig:njets}.
To perform clusterization with the \textit{direct} and \textit{fjcore} methods we selected 30, 60, 90 and 120 bins 
for $\yc$, the first number of bins being closer to experimental setups, while the last one is more typical 
for theoretical predictions.
In the \textit{transition} method $y_{\mathrm{ini}}$ was always set to the largest value of 
$\yc$ of the histogram, while $y_{\mathrm{stop}}$ was chosen to be the smallest. 
This way we ensured that the range of search for transition values coincides with the 
range of the histograms.

We summarize our results in Table \ref{tab:perf}.
\begin{table*}[ht]
\centering
\caption{Required time to perform clusterization of 1000 partonic events using the three different methods. 
Time values are shown is seconds. Various number of particles and bins were used to illustrate performance behavior.}
\label{tab:perf}
\begin{tabular}{cccccc}
\hline
Partons & Method & 30 bins & 60 bins & 90 bins & 120 bins \\
\hline
\hline
\multirow{3}{*}[-1pt]{5} 
& direct     & 0.103 s & 0.203 s & 0.285 s & 0.369 s \\ \noalign{\vspace{1pt}}\cline{2-6}\noalign{\vspace{2pt}}
& fjcore     & 0.167 s & 0.287 s & 0.406 s & 0.532 s \\ \noalign{\vspace{1pt}}\cline{2-6}\noalign{\vspace{2pt}}
& transition & 0.021 s & 0.023 s & 0.023 s & 0.023 s \\
\hline
\hline
\multirow{3}{*}[-1pt]{10} 
& direct     & 0.548 s & 1.062 s & 1.578 s & 2.091 s \\ \noalign{\vspace{1pt}}\cline{2-6}\noalign{\vspace{2pt}}
& fjcore     & 0.234 s & 0.421 s & 0.580 s & 0.777 s \\ \noalign{\vspace{1pt}}\cline{2-6}\noalign{\vspace{2pt}}
& transition & 0.296 s & 0.304 s & 0.299 s & 0.307 s \\
\hline
\hline
\multirow{3}{*}[-1pt]{15} 
& direct     & 1.635 s & 3.134 s & 4.532 s & 6.161 s \\ \noalign{\vspace{1pt}}\cline{2-6}\noalign{\vspace{2pt}}
& fjcore     & 0.313 s & 0.601 s & 0.887 s & 1.220 s \\ \noalign{\vspace{1pt}}\cline{2-6}\noalign{\vspace{2pt}}
& transition & 1.458 s & 1.740 s & 1.722 s & 1.746 s \\
\hline
\hline
\multirow{3}{*}[-1pt]{20} 
& direct     & 3.878 s & 6.930 s & 10.219 s & 13.548 s \\ \noalign{\vspace{1pt}}\cline{2-6}\noalign{\vspace{2pt}}
& fjcore     & 0.455 s & 0.896 s & 1.376 s & 1.927 s \\ \noalign{\vspace{1pt}}\cline{2-6}\noalign{\vspace{2pt}}
& transition & 5.357 s & 5.986 s & 5.924 s & 5.900 s \\
\hline
\end{tabular}
\end{table*}
The computations were performed on a simple everyday laptop.
We emphasize that our numbers in Table \ref{tab:perf} are shown just to illustrate 
the behavior of the new method compared to the usual one, it is not an exhaustive study 
on performance.
For example fluctuations in computational time were not taken into account.
Nevertheless Table \ref{tab:perf} still provides useful information about how the 
\textit{transition} method performs.

As we can see the timing of the \textit{direct} and the \textit{fjcore} methods scale with the 
number of bins, as one would expect it. 
The numbers indicate a linear relation.
The \textit{transition} method depends non-linearly on the number of particles, 
as more particles introduce more and more possible final jet configurations,
hence more transition values to compute.
This method also depends on the range of $\yc$ values.
Although a large number of particles would mean plenty of transition values,
many of them could fall outside of the range of interest, hence they would be not 
computed in the end.
For large number of partons the \textit{fjcore} method is the best, however 
there is a turnover at 10 partons, where the \textit{transition} method starts to take over and 
for 5 partons it clearly outperforms the other two methods, 
an order of magnitude speed up can be achieved.
It is even more obvious when the number of bins is larger.
Interestingly at low multiplicities \textit{fjcore} is the slowest, which is probably 
due to the complexity of the algorithm.

Table \ref{tab:perf} clearly shows that the \textit{transition} method can be used to improve 
the speed of clustering in the calculation of fixed order parton level distributions, like jet rates. 
The calculation of fixed order predictions typically involve only a small number of strongly 
interacting final state particles, but a large number of bins in order to produce smooth histogram curves.
In addition, millions of phase space points are generated, which all require clusterization, 
therefore faster methods are preferred.

We note that according to Table \ref{tab:perf}, the \textit{transition} method is always 
faster than the \textit{direct} method, when the number of bins is large and the multiplicity is moderate or small. 
As mentioned earlier both methods employ a 'naive' $\ordo{N^3}$ clusterization. 
Our new method does not depend on whether the clusterization is done via a 'naive' $\ordo{N^3}$ or 
an improved $\ordo{N \log N}$ algorithm, given that the value of $y_{ijk}^{max}$ can be tracked during repeated clustering. 
Hence we expect the integration of the \textit{transition} method into the \texttt{FastJet} framework to be possible. 

\section{Summary}
\label{sec:summary}
In this paper we defined transition values for the anti-$\kt$ algorithm
and we presented a way to compute them.
The knowledge of these values can speed up computations, which involve large number of variations of the 
$\yc$ jet parameter.
Our simple performance test shows that the new method could be applied best to improve performance significantly 
in the calculation of fixed order predictions for jet rates with the anti-$\kt$ algorithm, which 
might regain interest in the upcoming precision era and future electron-positron colliders. 
The definition of transition values could also serve as starting point for the development of new observables. 
Our method can be used both for the hadron and the $\epem$ collider version of the anti-$\kt$ algorithm, 
in fact for any version of the general inclusive $\kt$ algorithm.
Furthermore the new transition method can be combined either with the 'naive' $\ordo{N^3}$ or 
the improved $\ordo{N \log N}$ clusterization method, hence it is compatible with the \texttt{FastJet} framework. 

\subsection*{Acknowledgements}
We would like to thank Stefan Weinzierl for his useful comments on the manuscript.



\end{document}